# 4D Computational Ultrasound Imaging of Carotid Artery Flow

Yuyang Hu, Student member, IEEE, Michael Brown, Member, IEEE, Didem Dogan, Member, IEEE, Mahé Bulot, Maxime Cheppe, Guillaume Ferin, Member, IEEE, Geert Leus, Fellow, IEEE, Antonius F.W. van der Steen, Fellow, IEEE, Pieter Kruizinga, Johannes G. Bosch, Member, IEEE

*Abstract*— Computational ultrasound imaging (cUSi) with few elements and spatial field encoding can provide high-resolution volumetric B-mode imaging. In this work, we extend its application to 4D carotid artery (CA) flow imaging using a custom large-aperture 240-element matrix probe. We implemented a frequency band–based matched filtering strategy that balances resolution and contrast. The system's inherent imaging capabilities were evaluated and validated in flow phantom and human CA experiments. In the phantom study, 3D/4D power Doppler image and speckle-tracking analyses confirmed the system's ability to resolve flow structures and hemodynamics. In the human study, the CA bifurcation flow structure and its local pulsatile flow dynamics were successfully reconstructed. These results demonstrate the feasibility of using a large-footprint, few-element cUSi system for 4D CA flow assessment.

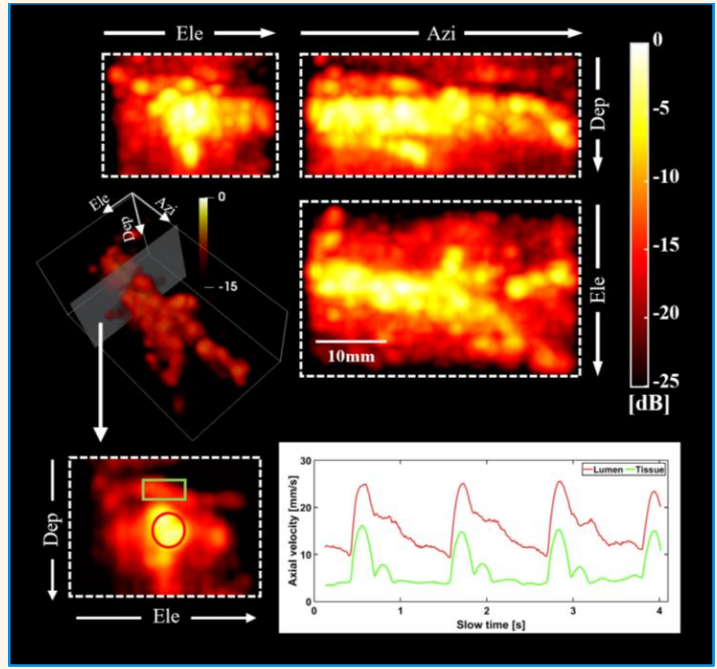

## I. INTRODUCTION

4D ultrasonography is considered a more convenient and powerful tool for carotid artery (CA) examination compared to conventional 2D ultrasonography. It enables more accurate quantification of stenosis and plaque volume [1]–[3], unhampered by scan-plane selection. Additionally, it allows for improved estimation of flow velocity and vessel wall motion/strain by incorporating displacement information in three dimensions [4]–[8]. These volumetric ultrasound measurements have traditionally been obtained through freehand scanning with a 2D linear array, which is highly operator-dependent [9], [10]. Mechanically swept transducers have been developed to alleviate this inconvenience [1], [11]. However, these systems remain limited to acquisition at low volume rates due to their beam-sweeping acquisition schemes.

A fully populated matrix array enables true 4D volumetric acquisition and supports plane-wave transmission for ultrafast imaging [12], [13]. For CA imaging, where a large aperture and high resolution are required, the spatial sampling limitation imposed by the Nyquist theorem would lead to a very high number of elements, and therefore, a very large number of independent channels. This results in considerable system complexity and cost on both the transducer and the receive hardware, posing significant challenges in terms of manufacturing feasibility and clinical accessibility. To address these limitations, strategies such as sparse arrays, row-column

This work was supported by the project TOUCAN (with project number 17208) of the OTP research programme which is financed by the Dutch Research Council (NWO).

This work involved human subjects. All ethical and experimental procedures and protocols were approved by the local medical ethics committee of Erasmus MC under application number MEC-2014-305 611.

Yuyang Hu, Antonius van der Steen and Johannes Bosch are with the Department of Cardiology, Erasmus MC University Medical Center, 3000 CA Rotterdam, The Netherlands (e-mail: y.hu@erasmusmc.nl).

Didem Dogan and Geert Leus are with the Department of Micro Electronics, Delft University of Technology, 2628 CJ Delft, The Netherlands.

Michael Brown and Pieter Kruizinga are with the Department of Neuroscience, Erasmus MC University Medical Center, 3000 CA Rotterdam, The Netherlands.

Mahé Bulot, Maxime Cheppe and Guillaume Ferin are with the Active Probe Group, Innovation Department, Vermon S.A., Tours, France。

*Highlights*

- The 4D carotid artery (CA) flow imaging ability of a custom computational ultrasound imaging (cUSi) system containing a large aperture 240-element matrix probe and a spatial encoding mask is evaluated.
- Our cUSi system is shown to be suitable for CA imaging by resolving the flow structures and the local pulsatile flow dynamics.
- This study demonstrates a potential pathway to a simple and cost-effective system for large-volume CA imaging and monitoring using the cUSi approach.

addressing, and sub-aperture beamforming have been proposed [14]–[18]. In our study, we investigate an alternative approach known as computational ultrasound imaging (cUSi) [19], [20]. This method combines spatial acoustic field encoding with model-based computational reconstruction, enabling high-resolution 4D ultrasound imaging with a substantially reduced number of channels, down to 1 [19] and 64 [20] in prior demonstrations, far below the Nyquist limits.

Based on this cUSi concept, we designed and fabricated a large-aperture ($40 \times 24$ mm²) matrix probe with only 240 elements for CA monitoring [21]. Coupled with a custom aberration mask for spatial encoding, the system's B-mode imaging capability was evaluated in phantom experiments, demonstrating volumetric image acquisition with promising resolution and contrast performance. In this paper, we further investigate the system's ability for 4D flow monitoring. Specifically, we propose an adapted matched filtering (MF) approach that operates in sub-frequency bands. The power Doppler image (PDI) and speckle-tracking methods is validated *in vitro*, and we present the first *in vivo* measurement of 4D carotid flow in a healthy volunteer. The rest of this paper is organized as follows: Section II introduces the overall cUSi scheme, provides a brief description of our probe and mask configuration, the reconstruction methods, experiments, and evaluation setups. Section III presents the comparison of different MF strategy choices and analyzes the *in vitro* and *in vivo* measurements. Section IV discusses the results and potential limitations, followed by conclusions in Section V.

## II. METHODS

### A. Computational ultrasound imaging

The goal of cUSi is to reduce the system hardware complexity for 3D imaging, i.e., needing fewer channels or sampling points within a large aperture, while maintaining acceptable B-mode and flow measurement quality for clinical diagnostic or monitoring use. In conventional imaging, reducing the number of measurements leads to undersampling artifacts and, therefore, decreased accuracy in reconstruction or estimation. In contrast, cUSi formulates this as a compressive imaging problem [19], [20], in which an encoding mask with a complex aberration pattern is used to spatially modulate the ultrasound field, enhancing the differences in pulse-echo responses (PERs) between spatial locations (often neighboring). Leveraging computational methods (model-based reconstruction) and prior knowledge of the imaging field, high-resolution images can be reconstructed from the undersampled channel signals.

We use a linear model to describe the signal formation of our cUSi system:

$$y_{(n_t, n_r)}(\omega, t) = \sum_{x=1}^{N_x} \sum_{y=1}^{N_y} \sum_{z=1}^{N_z} \left( A_{(n_t, n_r)}(\omega, x, y, z) \times x(x, y, z, t) \right). \quad (1)$$

In this model, $y$ represents the received channel signals, corresponding to different transmit $n_t$ and receive $n_r$ elements. For each transmit-receive pair $(n_t, n_r)$, every spatial location $(x, y, z)$ is associated with a PER, determined by the element positions and the spatial encoding of the acoustic field. The system matrix $A$ is constructed by assembling the PERs for all voxels within the region of interest (ROI), which consists of $(N_x \times N_y \times N_z)$ pixels. The vector $x$ contains the backscattering intensity at each spatial location in this linear model and represents the unknown ground truth we aim to reconstruct. For 4D flow estimation, $x$ is time-varying, with the fourth dimension, slow time, denoted by $t$. Furthermore, we express both the system matrix and the measured signals in the frequency domain, denoted by $\omega$, to facilitate the discussion of frequency-domain subsampling, reconstruction via convolution operations, and other frequency-domain techniques employed in this study.

### B. CUSi probe and mask

#### 1) System configurations

In short, the system consists of a 240-element matrix probe and an aberration mask for spatial encoding. As shown in Fig. 1a, the matrix probe comprises a $20 \times 12$ grid of square elements with a 2 mm pitch, resulting in a large aperture of $40 \times 24$ mm. The probe has unconventional frequency characteristics (Fig. 2a), with a center frequency of 6.6 MHz and a 6 dB bandwidth of 20%, and two additional peaks at 3.1 MHz and 9.3 MHz. When considering the 12 dB bandwidth, the total bandwidth reaches 116%.

The custom spatial encoding mask has a total thickness of 3 mm, composed of TPX and silicone rubber (Fig. 1a). A pattern of spatially varying layer thicknesses across the mask was used to introduce complex delay differences within a range of approximately 3 wavelengths, thereby achieving spatial acoustic field encoding. The mask remains detachable and replaceable, and is reproducibly aligned to the matrix array using a custom metal fixture.

More details can be found in [21], where this custom cUSi system is described and characterized for B-mode imaging.

*2) Imaging performance*

In the prior B-mode experiment reported there (Fig. 1b right), we acquired a volumetric image of size 40 × 24 × 50 mm using a CIRS phantom (Model 040GSE) combined with mask-based encoding. With the use of plane wave transmission, we achieved an effective volume rate of 500 Hz. The MF reconstruction of wire targets yielded an average lateral resolution of 1.08 mm and an axial resolution of 1.6 mm. A high-contrast inclusion region (specified to be >15 dB) was successfully reconstructed with a contrast ratio of 9.8 dB. When using the probe without the encoding mask (Fig. 1b left), a higher contrast ratio of 12 dB was obtained, but the lateral resolution degraded to 1.99 mm, while the axial resolution remained similar at 1.43 mm. Together with a prior *in silico* study of a similarly sized matrix array with fewer elements (5 × 5, 2 mm pitch) [22], we consider the cUSi system capable of supporting the flow and carotid artery estimation tasks in this study.

## C. Matched filtering reconstruction

For our cUSi system, we adopt model-based MF reconstruction methods [23]. This approach (eq. 2) is computationally efficient and well-suited to handle the complex aberrations introduced by the encoding mask, which are difficult to tackle using conventional methods such as DAS.

$$\hat{x}_{MF} = \sum_{n_t=1}^{N_t} \sum_{n_r=1}^{N_r} \left( A_{(n_t, n_r)}^H \times y_{(n_t, n_r)} \right). \quad (2)$$

This method evades directly inverting the system matrix $A$, which is infeasible due to its excessively large dimensions and ill-conditioning. By using the Hermitian matrix $A^H$ as a pseudoinverse, an approximate solution $\hat{x}_{MF}$ can be obtained. The system matrix was constructed using hydrophone measurements of element fields, which accurately capture the mask-induced delays as well as secondary effects such as internal reflections and mask attenuation. Note that the MF reconstruction can function as well without an aberration mask, which will result in decreased spatial separability but may have better clutter and SNR properties. We will compare results with and without the encoding mask.

In addition, we investigate MF using different radio-frequency (RF) frequency bands. In previous B-mode studies [21], we noticed that reconstruction quality was quite dependent on the choice of frequency components. Representation of the system matrix in the frequency domain makes this type of operation easily implementable. Particularly, frequencies lower than the center frequency benefit from less severe spatial undersampling, better acoustic SNR and penetration, especially useful with a mask. This approach is allowed by the relatively good sensitivity of our transducer around 3.3 MHz (Fig. 2a).

## D. Experiments

We conducted *in vitro* experiments using a flow phantom (CIRS ATS 524) and *in vivo* measurements in healthy volunteers to evaluate the performance of our reconstruction methods in the case of flow.

In the *in vitro* setup, the phantom was connected to a peristaltic pump (Verder Peristaltic Pump, 2006), which

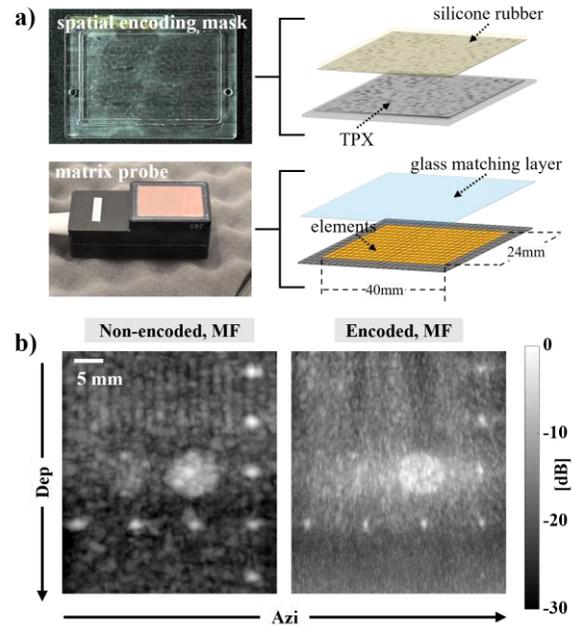

Fig. 1 a) Overview of the proposed cUSi system as described in [21]. Top image: the spatial encoding mask, composed of a patterned TPX slice and a layer of silicone rubber. Bottom: the 240-element matrix probe, featuring an active aperture of 40 × 24 mm², formed by a 20 × 12 grid of elements, and incorporating a glass matching layer. b) MIPs of 3D B-mode reconstruction acquired using the cUSi system with the matrix probe coupled without (left) and with (right) the encoding mask. Imaging was performed using 16 angle plane wave transmission and reconstructed via MF method.

circulated blood-mimicking fluid (CIRS Model 769DF) through a 6 mm diameter channel at a peak velocity of 15 cm/s. The matrix probe was excited with 4-cycle tone bursts at 3 MHz and 6.25 MHz. These bursts were directly concatenated in transmission, enabling sub-frequency band reconstruction comparisons within the same acquisition volumes. Plane wave transmissions were implemented using a 4×4 angular steering scheme, with uniformly distributed angles over ±12° in azimuth and ±7° in elevation, four angles per direction. The pulse repetition frequency was 10 kHz, resulting in an effective volume rate of 625 Hz. The probe was positioned along the phantom's long axis, aligned with the flow direction, and a gel pad was inserted to introduce an approximate 10° Doppler angle. RF data were acquired from a depth range of 5–55 mm over a 2-second duration.

In the *in vivo* measurements, a 6.25 MHz, 4-cycle tone burst was used, employing the same 16-angle plane wave transmission scheme as in the *in vitro* setup but with a 500 Hz PRF. The matrix probe was positioned over the CA with a stand-off gel pad of 20 mm thick, and RF data were acquired from a depth range of 10–50 mm for 4 seconds. For both *in vitro* and *in vivo* studies, acquisitions were performed with and without the encoding mask.

## E. Evaluations

*1) Inherent imaging capabilities of frequency band-based MF: correlation map analysis*

First, we assessed our cUSi system's performance across different frequency bands. In our previous study [21], we

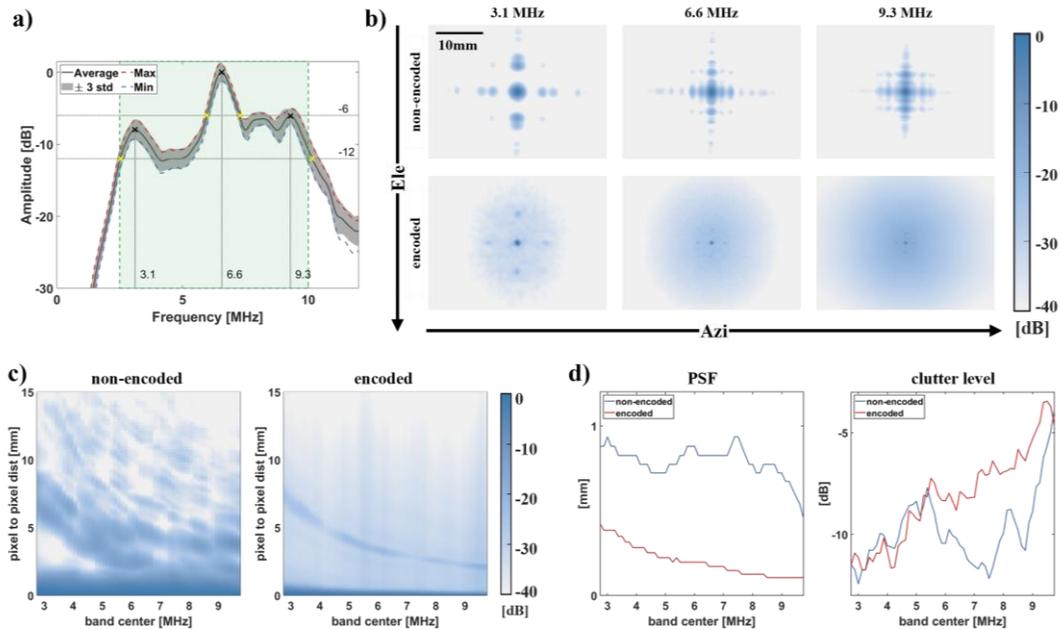

Fig. 2 a) Frequency response of the cUSi matrix probe. The green box indicates the frequency range used for correlation map analysis, and the yellow crosses mark the boundaries for the −6 dB and −12 dB bandwidth calculations. b) Averaged PER correlation maps at 25 mm depth, showing the signal similarity between each pixel and its surrounding pixels. From left to right, the center frequencies of the bands increase from 3.1, 6.6, to 9.3 MHz. Top row: without encoding; bottom row: with encoding. c) 2D heatmaps showing PER correlation magnitude as a function of pixel-to-pixel distance and band center frequency. Left: without encoding; right: with encoding. d) Expected PSF width (left) and clutter level (right), both derived from the correlation analysis in (c). As demonstrated in this figure, encoding consistently improves PSF sharpness across all frequency bands. Higher frequencies result in narrower PSFs but also exhibit increased clutter levels, particularly when encoding is applied.

observed that the choice of frequency band has a significant impact on image quality. To verify this, we analyzed the correlation maps of PERs across frequency bands. Since these maps are derived directly from the system matrix, they should objectively reflect the system's inherent imaging capabilities, independent of specific imaging conditions. Based on the transducer's frequency response characteristics with the three major response peaks, we selected the 2.5–10 MHz range for analysis(Fig. 2a, green window), using a sliding window of 0.5 MHz bandwidth (band centers 2.75-9.75 MHz, 0.125 MHz step).

Specifically, a 40 × 24 mm lateral plane at a depth of 25 mm was selected for evaluation, with a spatial sampling step of 0.2 mm. For each pixel in this plane, we computed the correlation of its PER with that of all other surrounding pixels. The averaged correlation map of all pixels can be interpreted as representing the system's lateral PSF and clutter sensitivity at that depth[20]. Given the point-symmetric distribution of patterns around the center, correlation values were aggregated over all angles and plotted as a function of distance, yielding a 1D polar projection of the PSF. To preserve features such as secondary peaks, we used the average of the top 5% of correlation values at each distance rather than taking a simple mean. Finally, we normalized each frequency band independently and calculated the full width at half maximum (FWHM) of the polar projection as an estimate of PSF width, and defined the residual energy beyond the FWHM as the clutter level. This allowed us to quantitatively assess the system's inherent imaging capabilities based on PER correlation analysis.

### 2) Flow phantom validation

We compared flow phantom MF reconstructions from two representative frequency bands, using both encoded and non-encoded cases. RF data were first slow-time high-pass filtered at 30 Hz to obtain the Doppler signal of interest. This signal was then filtered in fast time to obtain two sub-bands: 2.4–4.0 MHz (low-band) and 5.4–7.0 MHz (high-band), each containing approximately 200 frequency samples. These were then used for PDI reconstruction [24].

The 4D PDIs were averaged over 1.5 seconds of acquisition containing relatively constant flow (~900 reconstructed volumes). A spatial mean filter with a 0.5 mm cubic kernel was applied to improve visualization. The contrast ratio between the flow lumen and a background region at the same depth was quantified as a performance metric to compare the four reconstruction cases.

To demonstrate the cUSi system's spatiotemporal resolution capability, we applied a simplified 1D speckle tracking approach as proof of concept [25]. The 4D PDI reconstructions were first projected onto the azimuth–slow-time plane by maximum-intensity projection (MIP) along depth and elevation and then tracked along slow-time to estimate flow velocity along the longitudinal direction of the channel. The template window size was empirically set to 5 mm (typically covering 3–4 speckle trajectories), with a 7 mm search window and a step size of 0.2 mm. To extract reliable and non-redundant velocity estimates, only tracking results meeting the following criteria were retained for statistical quantification: (1) the estimation was located at the center of a flow speckle, as identified from local maxima in the projection; (2) the correlation coefficients for both the current estimation and its predicted positions over the next two time steps exceeded 80%.

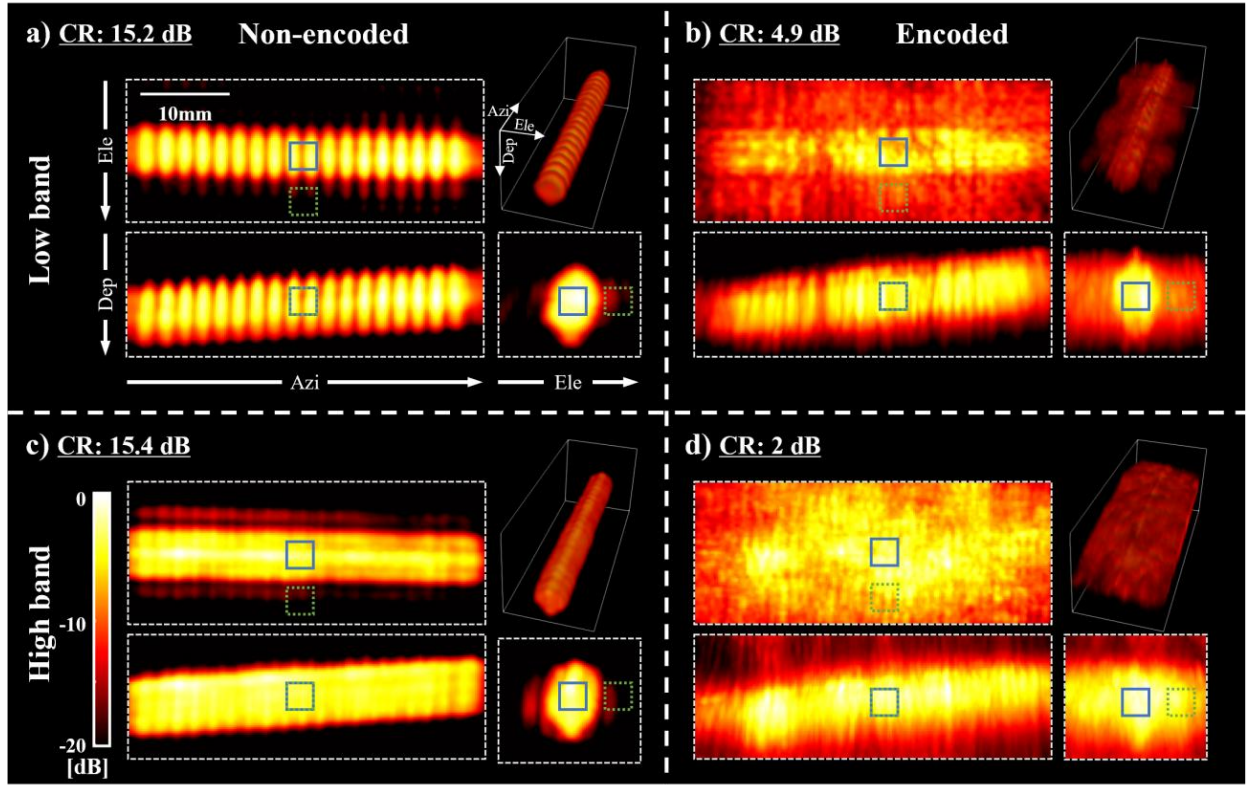

Fig. 3 Volumetric PDI reconstructions of the flow phantom under non-encoded (left) and encoded (right) conditions, shown for two frequency bands: low band (top row) and high band (bottom row). Each subfigure includes MIPs in three orthogonal dimensions, with a 3D rendering displayed in the top-right corner (shown at 15 dB). CR is quantified using two cubic ROIs inside and outside the channel at the same depth, marked by blue and green boxes in each MIP.

For velocity comparisons, conventional 2D scans and spectral Doppler acquired from the flow phantom using a clinical ultrasound system (Zonare ZS3, Mindray, China) with a L14-5w linear array transducer were used to confirm the peak velocity of 15 cm/s (supplemental Fig. S1).

3) In vivo validation

In vivo validation was performed using non-encoded high-band MF, since we found that its contrast performance was better than that of the encoded low-band. A spatial average filter with a 2 mm cubic kernel was applied to the reconstructed volumes for noise suppression.

The 4D PDI was first reconstructed from non-clutter filtered data. By checking the total power variation over slow time, the different cardiac cycles were identified. A high-pass wall filter with a 100 Hz cutoff was then applied, and volumes corresponding to the diastolic phase, from the notch stage to the following systole, were extracted across the three cycles (~1000 volumes in total). These were then averaged over slow time to generate the final 3D PDI.

Subsequently, a cross-sectional slice at the common carotid artery (CCA) region was selected, covering 27 × 17 mm in the elevation–depth plane. For each pixel in this plane, spectrograms were computed using a 128-sample (~0.25 s) temporal window, sliding with a step size of 4 samples [26].

III. RESULTS

A. Correlation map analyses

Fig. 2 presents the PER correlation analysis across different frequency bands. As shown in Fig. 2b (top row), in the non-encoded acoustic field, each pixel exhibits a strong correlation with its immediate neighbors, resulting in a strong central peak. Beyond this main peak, multiple secondary peaks are observed, regularly distributed along the azimuth, elevation, and diagonal directions of the probe field. These features are believed to correspond to the main lobe and grating lobes of the PSF in the reconstructed image. As the frequency increases, the grating lobes appear at smaller angles relative to the main lobe, thereby moving closer to the central peak in the correlation map (Fig. 2c).

With the use of encoding mask (Fig. 2b, bottom row and Fig. 2c, right), the amplitude of the grating lobes is significantly suppressed, and the diagonally distributed correlation peaks become invisible. Additionally, the background region outside the central peak exhibits increased correlation levels, indicating a higher clutter level. Most importantly, compared to the non-encoded case, the encoded field consistently exhibits a sharper PSF peak across all frequency bands, suggesting improved spatial resolution.

Fig. 2d quantifies the behavior of the PSF and clutter level as functions of frequency. As visible in Fig. 2c, the encoded case yields consistently sharper PSFs than the non-encoded

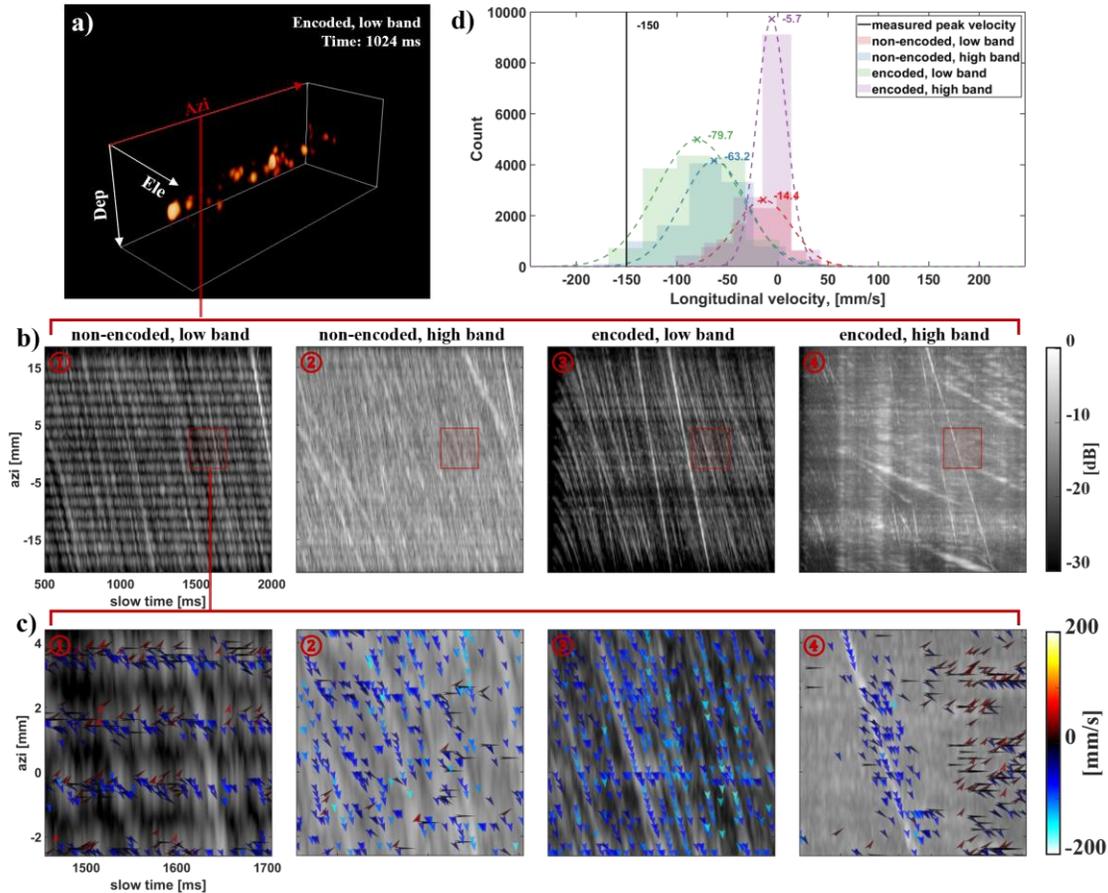

Fig. 4 a) Reconstructed moving speckle pattern in the flow phantom, shown for the encoded, low-band case at the 1024 ms timepoint (see Supplementary movie M2 for the full sequence.) b) MIP along the azimuth direction over slow time. Each trace corresponds to the longitudinal movement of a speckle along the flow channel. From left to right: non-encoded low and high frequency band reconstructions; encoded low and high band. c) 1D speckle tracking results within the zoomed-in red ROIs marked in (b), shown as arrows aligning with the orientation of the speckle traces. The arrow colors indicate local displacement magnitude (velocity). d) Histogram of estimated velocity for all four cases, fitted with Gaussian distributions (dashed line).

case. In terms of clutter, the non-encoded case demonstrates a lower level at higher frequencies. Furthermore, the encoded results exhibit smooth and near-linear frequency dependence in both PSF sharpness and clutter level, whereas the non-encoded results appear more fragmented and exhibit irregular frequency dependence.

### B. Flow phantom PDI reconstructions

Fig. 3 shows the PDI reconstructions of the flow phantom for different frequency bands, averaged over a 1.5-second acquisition. Visually, the encoded low-band (Fig. 3b) and the non-encoded high-band (Fig. 3c) yield superior reconstruction quality compared to the other two cases (Fig. 3a and 3d). The non-encoded high-band case (Fig. 3c) provides the most complete flow profile (CR = 15.4 dB), with clear and continuous lumen boundaries. Slight artifacts are observed along the elevation direction (i.e., to the left and right of the main channel in the channel cross-section), likely due to grating lobes. The encoded low-band case (Fig. 3b) also presents a relatively natural and coherent flow structure, although increased clutter is visible compared to Fig. 3c (CR = 4.9 dB), and the channel's uniform flow is partially obscured. The non-encoded low-band case (Fig. 3a) exhibits good contrast performance (CR = 15.2 dB), but the reconstructed flow pattern displays periodic intensity variations along the lumen. Lastly, the encoded high-band case (Fig. 3d) suffers from excessive clutter (CR = 2 dB), which severely degrades lateral discrimination of the vessel limits. As a result, the vessel profile is only faintly discernible along the depth axis and nearly unresolvable in the elevation direction.

### C. Speckle-based velocity analysis

To further investigate the time-resolved performance of our system, Fig. 4 presents speckle-based flow estimation results in the flow phantom along the slow-time dimension. As shown in Fig. 4a, the cUSi system successfully reconstructs clearly identifiable speckles within a single volume. The 4D rendering of PDI is provided in supplementary movies M1 and M2. In these videos, these speckles can be observed flowing continuously through the entire channel, visually confirming temporal coherence. The azimuth-slow-time MIPs in Fig. 4b depict the spatiotemporal behavior of speckles. Each trace represents a moving speckle, with the slope of the trace indicating its local velocity. Note that the two non-encoded cases (Fig. 4 b-1 and b-2) are derived from the same acquisition, and both encoded cases (Fig. 4 b-3 and b-4) from a second acquisition. Despite the suboptimal PDI performance in Fig. 3b, the encoded low band case (Fig. 4b-3) yields the most numerous, sharp, and distinguishable speckle traces. The second-best result is observed in the non-encoded high band (Fig. 4b-2), where many traces are present but appear less sharp and less clear. We attribute this to reduced contrast at high frequencies and lower resolution in the absence of encoding. The non-encoded low band (Fig. 4b-1) shows relatively high contrast but suffers from discontinuous, step-like speckle traces, more visible in the zoom-in of Fig. 4c-1. In addition, striping artifacts in intensity are observed along the azimuth direction, a phenomenon also present in the PDI reconstruction of Fig. 3a. Finally, the encoded high band (Fig. 4b-4) produces only a few visible speckle traces that protrude above the background.

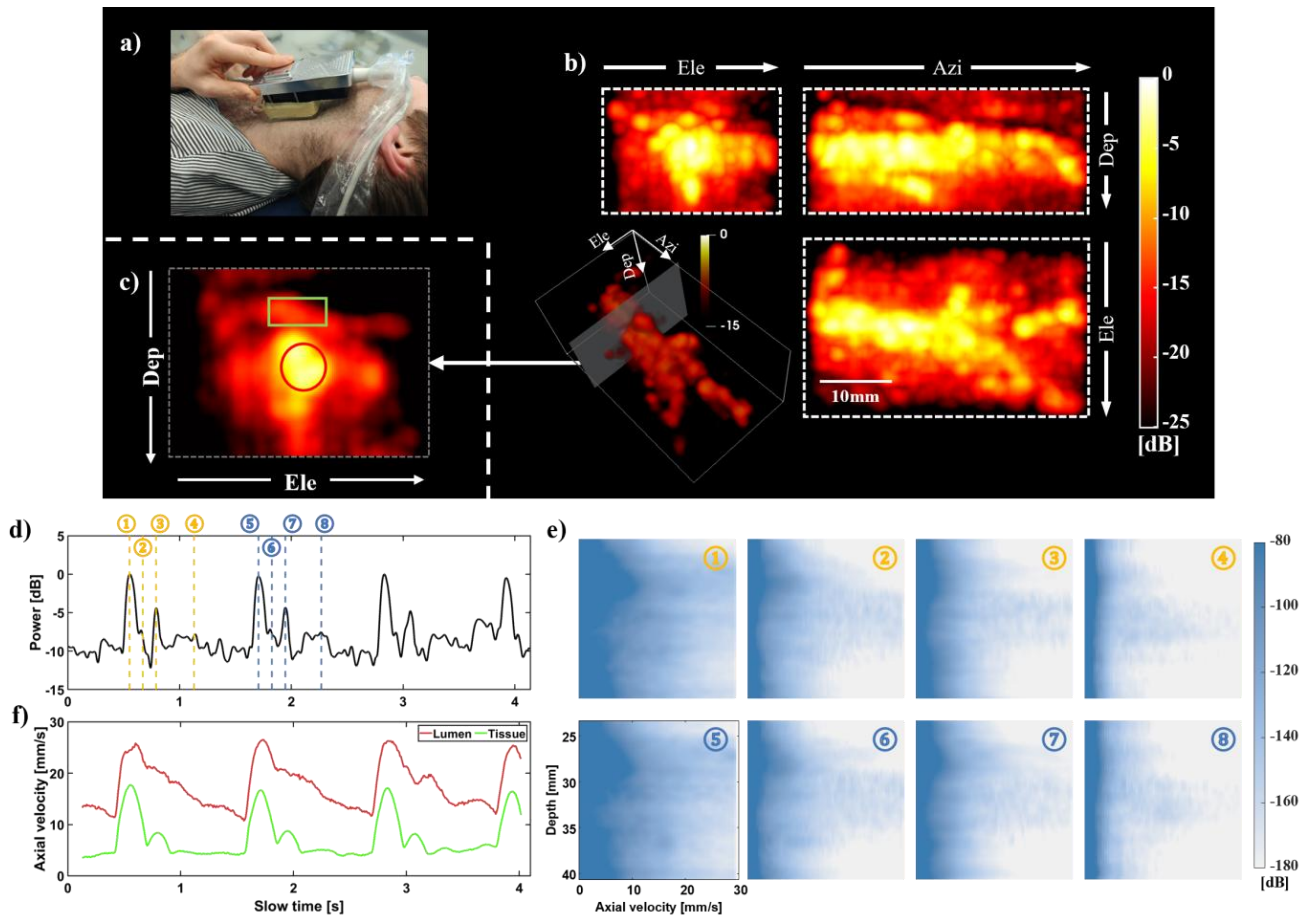

Fig. 5 In vivo validation of the CA from a healthy volunteer, measured and reconstructed using non-encoded high-band MF. a) Photograph of the matrix probe placed on the volunteer's neck. b) 3D PDI results displayed as MIPs along three axes and a 3D rendered view. c) Azimuth cross-sectional view of the PDI at CCA region. The red circle and green box indicate ROIs within the vessel lumen and surrounding tissue, respectively. d) Total Doppler power variation over slow time for the entire cross-section, before clutter filtering. e) Averaged spectrograms by depth (prior to clutter filtering) for multiple cardiac phases, corresponding to the time markers indicated in d. f) Average velocity in lumen (green box in c) and tissue (red circle in c) regions, computed after applying the high-pass clutter filter.

These differences in image quality further impact the performance of 1D speckle tracking, as shown in Fig. 4c. The encoded low-band case (Fig. 4c-3) exhibits the highest number of valid traces and demonstrates strong consistency in displacement estimation. In the non-encoded high-band (Fig. 4c-2), some erroneous estimations are observed, including some apparent displacements against the actual flow direction. The non-encoded low-band (Fig. 4c-1) produces many incorrect estimations as well, likely due to the step-like trace patterns and striping artifacts. In the encoded high-band (Fig. 4c-4), only a small number of traces can be identified due to the high clutter level. These tracking outcomes are further analyzed by their velocity distributions shown in Fig. 4d. The encoded low-band case yields a distribution centered around 10 cm/s, which is closer to the actual peak velocity of 15 cm/s compared to the other cases. The distributions from the remaining cases are consistent with the qualitative observations discussed above.

### D. In vivo validation

Fig. 5 presents the PDI reconstruction results from an *in vivo* validation in a healthy volunteer. A ROI measuring 42 × 26 × 18 mm was reconstructed at a depth of 32 mm. The non-encoded high-band MF successfully reconstructs identifiable CA flow structures. As shown in Fig. 5b, the geometry of the carotid bifurcation is clearly visualized, including the upstream CCA and the downstream internal carotid artery (ICA) and external carotid artery (ECA) branches. A rotating rendered 3D PDI is shown in supplementary video M3. Further analysis on a reconstructed cross-sectional slice (azimuth ≈ 12 mm) (Fig. 5c) reveals detailed temporal dynamics. From the total instantaneous power prior to wall filtering, we can observe three distinct and complete cardiac cycles (Fig. 5d).

The depth-resolved Doppler spectrograms in Fig. 5e further show how the velocity-depth distributions vary over four cardiac stages (left to right) and are similar in two consecutive heartbeats (top vs. bottom). The depth-resolved spectrograms across slow time are provided in the supplementary video M4. We first notice that the spectrograms at all cardiac stages exhibit strong low-velocity components. Across the cardiac cycle, these components show characteristic changes: a sudden shift from lower to higher velocities during systole; a subsequent reduction; a modest resurgence during the notch phase; and finally, a steady decline approaching zero during diastole. Notably, these fluctuations present two pronounced velocity peaks across depth, which we interpret as corresponding motion from the vessel walls, while other regions above and below likely represent surrounding tissue. Interestingly, within the region lying between the two vessel walls, we consistently observe high-velocity components, forming a roughly parabolic contour. We attribute this velocity pattern, which is weaker in intensity than the tissue velocities, to blood flow velocity within the lumen. The average (axial) blood velocity per pixel was estimated from the spectrogram after empirically discarding frequency components with power above -105 dB to suppress the influence of low-velocity clutter motion.

Fig. 5f shows the velocity over time in both the lumen and tissue/wall regions, as part of a further investigation into Doppler signal components. These curves were averaged from clutter-filtered pixel velocities derived from spectrograms, using the red circle (lumen) and green box (tissue/wall) ROIs marked in Fig. 5c. The lumen region (red) exhibits a significantly higher mean velocity than the tissue region (green). Moreover, a marked flow velocity increase occurs following each systolic wall motion, consistent with expected behavior [27].

## IV. Discussion

In this work, we investigated the performance of a customized cUSi system for CA flow monitoring, employing a 240-element large-aperture matrix probe with and without an encoding mask. We explored the effects of the mask and using different frequency components in the RF signal for our MF reconstruction to come to the most useful reconstruction of the flow. We first compared the inherent imaging capabilities of the cUSi system for RF frequencies between 2.5-10 MHz (Fig. 2) by exploring the PER correlation maps of pixels and their neighborhood, directly from the constructed system matrices, both with and without spatial encoding. The analysis revealed that markedly different performance is obtained for different chosen frequency bands. We tested the system's performance in a flow phantom experiment. 4D PDIs were reconstructed across different RF frequency bands. Time averaging into 3D PDIs successfully visualized the flow structures within the phantom (Fig. 3). Furthermore, 4D PDI enabled the visualization of clearly distinguishable moving speckles within the flow region. The quality of these speckles was sufficient to support speckle–tracking–based flow estimation. Finally, in an *in vivo* test on a healthy volunteer, we successfully reconstructed the PDI of the CA bifurcation, with visible flow in the CCA, ECA, and ICA. Spectrogram analysis revealed distinct temporal power variations between the lumen and surrounding tissue, separating tissue and flow components.

### A. Frequency band and encoding

Regarding the frequency-dependent behavior observed in Fig. 2, the encoded field exhibits a trend in both expected PSF and clutter level, which is proportional to frequency (Fig. 2d, red), whereas this trend appears less consistent in the non-encoded field. We attribute this to the combined effects of spatial sampling and acoustic encoding, which interact differently across frequency bands. As frequency increases, the effective aperture sampling improves and the main lobe becomes narrower, potentially enhancing lateral resolution. However, due to the relatively larger pitch (compared to the shortening wavelength), grating lobes shift closer to the main lobe and may even overlap with it (as shown in Fig. 2b, middle and right), resulting in more complex PSF profiles. With encoding, the spatial periodicity of the array is disrupted, effectively suppressing grating lobes and redistributing acoustic interference more broadly. In addition, as the wavelength shortens with increasing frequency, a fixed encoding mask introduces greater relative delays (in normalized wavenumbers), increasing spatial decorrelation between adjacent pixels. This leads to further PSF sharpening at higher frequencies, albeit at the cost of increased clutter levels. Together, these mechanisms explain the consistent PSF improvement and rising clutter level observed in the encoded cases, and conversely, the less predictable behavior in the non-encoded cases.

Fig. 3 presents the *in vitro* validation of our system matrix–based analysis, where the non-encoded high-band and the encoded low-band MF reconstructions of the flow phantom demonstrated superior performance in 3D PDI compared to the other two cases. We conclude that these two configurations achieve a favorable trade-off between contrast and PSF characteristics, albeit via different mechanisms. For the encoded cases (Fig. 3b, d), applying spatial encoding increases element field overlap [21] and suppresses grating lobes. Although the encoding mask tends to reduce the signal SNR and elevate the clutter level, selecting the lower frequency band mitigates these losses. Together, these factors contribute to an improved overall reconstruction quality in the encoded low-band scenario. Without encoding, CR is less problematic, but grating lobes start to affect the reconstructions. In Fig. 2b, correlation peaks representing strong grating lobes are seen in the non-encoded case. At higher frequencies, these lobes overlap with the main lobe, smoothening intensity variations. As a result, the non-encoded high-frequency MF produces a more continuous and uniform flow structure than at lower frequencies (Fig. 3c vs. a).

### B. Speckle tracking potential

Temporal speckle analysis, as shown in Fig. 4, demonstrates that our system is capable of delivering temporally and spatially resolved measurements, especially for encoded low-band and non-encoded high-band. Speckle tracking was applied in 1D as a proof of concept. Based on visual inspection of the supplementary movies M1 and M2, we anticipate that the speckle tracking could also perform effectively in a full 3D/4D scenario, although this was not attempted in this study. The encoded low-band reconstruction produced results closest to the commercial device's measured peak velocities. However, this encoded low-band reconstruction still underestimated the flow velocity as measured by the commercial 2D linear array system near the center of the lumen. This deviation is explainable because the 1D MIP favors the brightest particle across the lumen cross-section, which may move at a velocity lower than the maximum.

### C. Alternative reconstruction approaches

Regarding the employed reconstruction method, this paper focused particularly on the band-based MF. This technique allows some tuning of the cUSi imaging, enabling a trade-off for better resolution or contrast, while reducing the computational burden by processing only a subset of the frequencies. Importantly, such adjustments can be easily implemented by system matrix manipulation in a model-based reconstruction framework that is formulated in the frequency domain. Another method that modifies the system matrix is the phase-only MF proposed by Brown et al. [20], which aims to mitigate artifacts arising from sensitivity to field pressure distributions, an issue inherent to conventional MF, as discussed in [28]. In this approach, only the phase of the system matrix is utilized during reconstruction, effectively introducing a magnitude-based normalization into the MF process. However, the successful application of phase-only MF requires precise clutter filtering to isolate the flow-related phase information, a step that is challenging to automate in our CA application. We include one example case in the supplementary materials (Fig. S2) to demonstrate the feasibility of the phase-only MF concept, but further investigation is certainly warranted. Furthermore, in our previous B-mode study, we demonstrated that iterative methods such as LSQR can further improve reconstruction quality [21], [29]. However, in the current study, we did not explore such approaches due to their significantly higher computational demands, especially in the context of flow imaging, where extremely long time-resolved ground truth data must be processed. Future work will explore iterative reconstruction for cUSi flow imaging.

### D. Future research

This study serves as a preliminary proof-of-concept, with *in vivo* validation performed using the non-encoded high-band MF reconstruction, limited to 3D PDI and time-resolved spectrogram analysis. For future clinical translation, several aspects require further development.

Firstly, the system's sensitivity and contrast need improvement. *In vivo* imaging poses additional challenges due to tissue-induced aberration and attenuation, which degrade reconstruction quality. Potential strategies include using lower transmission frequencies, thinner encoding masks to preserve SNR, and applying contrast agents to enhance SNR and enable methods such as echo-particle image velocimetry as a potential follow-up [30].

Secondly, the current volume rate is insufficient to fully capture the rapid flow dynamics in the CA, primarily due to the empirically selected 16-angle plane wave transmitting scheme, which aims to preserve reconstruction quality. Future work should explore the possibility of accurately extracting flow information with fewer transmissions. Additionally, sub-band MF with a narrower bandwidth can be used to reduce data throughput and improve frame rate [31].

Thirdly, the current probe exhibits unconventional frequency characteristics due to its acoustic stack design [21]. Probes with a broader and flatter frequency response may enable more effective utilization of different frequency components in future work.

Lastly, although the large aperture offers a wide imaging field, the probe and mask fixture's relatively bulky structure may hinder usability. In practice, we found it difficult to achieve proper coupling in anatomical regions such as the submandibular neck. A redesign of the probe housing or mounting fixture could improve clinical usability.

Ultimately, the proposed cUSi framework may lend itself well to wearable CA applications [32], offering enhanced image quality and monitoring with reduced hardware complexity.

## V. CONCLUSION

In this paper, we demonstrated the flow detection capability of a cUSi matrix array system aiming at CA imaging. To improve reconstruction quality in flow measurements, where relatively high PRFs are desired and thus the number of observations should be limited (16 angles, 240 channels), we employed frequency band–based MF to achieve an acceptable balance between contrast and resolution. In the flow phantom studies, both the non-encoded high-band and encoded low-band MF demonstrated relatively good reconstruction performance. The system's ability to extract hemodynamic information was demonstrated through volumetric PDI and speckle-based analysis in a flow phantom. Finally, *in vivo* validation successfully reconstructed time-resolved CA flow structures, including the CA bifurcation structure and flow dynamics, confirming the feasibility of using cUSi for 4D CA imaging.

## ACKNOWLEDGMENT

We thank Lucy Wei and Jason Voorneveld (Cardiology and Neuroscience, Erasmus MC) for advice on Verasonics acquisition and flow experiments; Robert Beurskens and Gerard van Burken (both at Cardiology, Erasmus MC) for valuable assistance with hardware; Stein Beekenkamp and Geert Springeling (Medical Instrumentation, Erasmus MC) for fabricating the encoding mask and experiments setups.

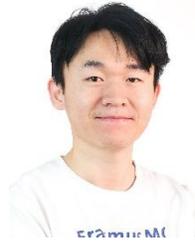

**Yuyang (FOX) Hu** (Student member, IEEE) studied and earned his bachelor's and master's degrees in biomedical engineering from Shenzhen University from 2011 to 2019. He is currently pursuing his Ph.D. degree in the Thoraxcenter Biomedical Engineering group, Department of Cardiology, Erasmus Medical Center, Rotterdam, the Netherlands.

His research is about applying the computational ultrasound imaging for the carotid artery monitoring with low number of sensors.

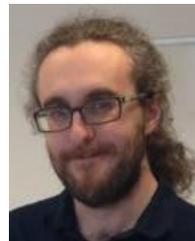

**Michael Brown** (Member, IEEE) got his MSc in physics from the University of Bristol in 2012, MRes in Medical and Biomedical imaging, and PHD from UCL in 2014 and 2018, respectively. He is currently a postdoctoral research fellow in the Department of Medical Physics and Biomedical Engineering at UCL and the Department of Neuroscience at Erasmus MC. His research interests are methods for ultrasound wavefront shaping and computational and functional imaging.

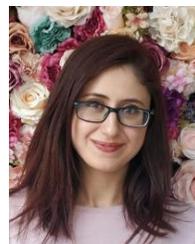

**Didem Dogan** (Member, IEEE) received the M.Sc. degree in electrical engineering in 2020 from Middle East Technical University, Ankara, Turkiye. She is currently pursuing the Ph.D. degree with the Signal Processing Systems department of the Delft university of Technology, Delft, Netherlands. Her research interests include the general area of signal processing, with an emphasis in computational imaging and particularly the ultrasound imaging.

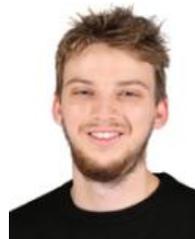

**Mahe Bulot** studied and earned his master's degree in acoustic and vibration engineering in ENSIM school based in Le Mans. He is currently part of Active Probe department in Vermon as acoustic design engineer.

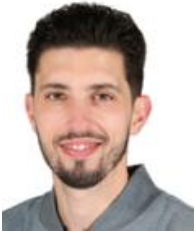

**Maxime Cheppe** obtained his Professional Bachelor's degree in Biomedical Microtechnology Engineering in 2005 from the Tours Polytechnic school. He is a mechanical designer in the Active Probe department at Vermon SA (Tours) since 2013.

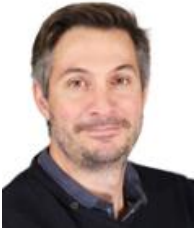

**Guillaume Ferin** is an enthusiastic engineer with a PhD in ultrasound technology for medical imaging from Université François Rabelais (2006). Since 2002, he has been with VERMON in Tours, France, where he focuses on the design and manufacturing of ultrasound transducers. His career includes a period at Vizyontech Imaging (2012-2014) developing advanced 3D ultrasound solutions for early breast cancer detection before returning to VERMON's innovation department. For the last six years, he has led the Active Probes service at VERMON, overseeing the design of innovative 2D and 3D ultrasound devices with embedded electronics.

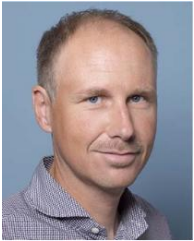

**Geert Leus** (Fellow, IEEE) received the M.Sc. and Ph.D. degrees in electrical engineering from the KU Leuven, Leuven, Belgium, in June 1996 and May 2000, respectively. He is currently a Full Professor with the Faculty of Electrical Engineering, Mathematics and Computer Science, Delft University of Technology, Delft, The Netherlands. He was the recipient the 2021 EURASIP Individual Technical Achievement Award, a 2005 IEEE Signal Processing Society Best Paper Award, and a 2002 IEEE Signal Processing Society Young Author Best Paper Award. He was also Member-at-Large of the Board of Governors of the IEEE Signal Processing Society, Chair of the IEEE Signal Processing for Communications and Networking Technical Committee, Chair of the EURASIP Technical Area Committee on Signal Processing for Multisensor Systems, Editor in Chief of the EURASIP Journal on Advances in Signal Processing, and Editor in Chief of EURASIP Signal Processing. He is a Fellow of EURASIP.

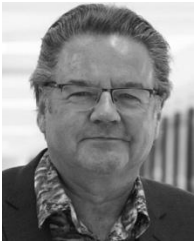

.**Antonius F. W. van der Steen** (Fellow, IEEE) received the M.Sc. degree in applied physics from Technical University Delft, Delft, The Netherlands, in 1989, and the Ph.D. degree in medical science from Catholic University Nijmegen, Nijmegen, The Netherlands, in 1994. He is currently the Head of biomedical engineering with the Thorax Center, Erasmus MC, Rotterdam, The Netherlands. He is an expert in ultrasound, cardiovascular imaging, and cardiovascular biomechanics. He is a fellow of the European Society of Cardiology, member of the Netherlands Academy of Technology (AcTI) and Board Member of the Royal Netherlands Academy of Sciences (KNAW). He was a recipient of the Simon Stevin Master Award and the NWO PIONIER Award in Technical Sciences.

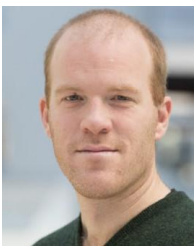

**Pieter Kruizinga** received the Ph.D. degree from Erasmus MC, Rotterdam, The Netherlands, in 2015.
In 2018, he joined the Neuroscience Department, Erasmus MC, where he leads the imaging research with the Center for Ultrasound and Brain-Imaging Erasmus MC (CUBE). His current research focuses on computational ultrasound imaging and functional ultrasound imaging of the brain.

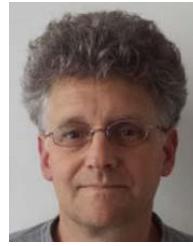

**Johan G. Bosch** (Member, IEEE) received the M.Sc. degree in electrical engineering from the Eindhoven University of Technology, Eindhoven, The Netherlands, in 1985, and the Ph.D. degree from the Leiden University Medical Center, Leiden, The Netherlands, in 2006.
He is currently an Associate Professor and a Staff Member with Thoraxcenter Biomedical Engineering, Department of Cardiology, Erasmus University Medical Center, Rotterdam, The Netherlands. His research interests include 2D and 3D echocardiographic image formation and processing, transducer development, and novel ultrasound techniques for image formation and functional imaging.

# Supplementary Material

# 4D Computational Ultrasound Imaging of Carotid Artery Flow

Yuyang Hu[1]*, Michael Brown[2], Didem Dogan[3], Mahé Bulot[4], Maxime Cheppe[4], Guillaume Ferin[4], Geert Leus[3], Antonius F.W. van der Steen[1], Pieter Kruizinga[2], Johannes G. Bosch[1]

[1] Department of Cardiology, Erasmus MC University Medical Center, 3000 CA Rotterdam, The Netherlands

[2] Department of Neuroscience, Erasmus MC University Medical Center, 3000 CA Rotterdam, The Netherlands.

[3] Department of Micro Electronics, Delft University of Technology, 2628 CJ Delft, The Netherlands

[4] Active Probe Group, Innovation Department, Vermon SA, Tours, France

*Corresponding author, E-mail address: y.hu@erasmusmc.nl (Y. Hu).

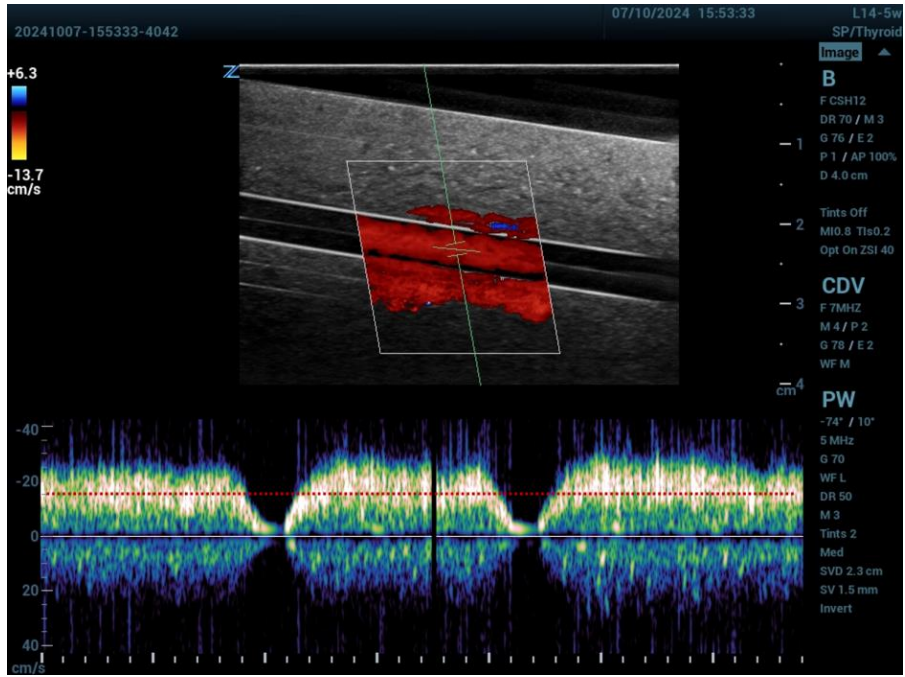

Fig. S1 Flow measurement in the phantom experiment using the Zonare ZS3 system with an L14-5w linear array transducer. Color Doppler velocity is overlaid on the B-mode image. Pulsed wave (PW) Doppler was measured at the center of the channel lumen. The results show that during peristaltic pump activation, the flow velocity in the channel remained around 15 cm/s. This measurement was used as a reference for the in vitro experiments.

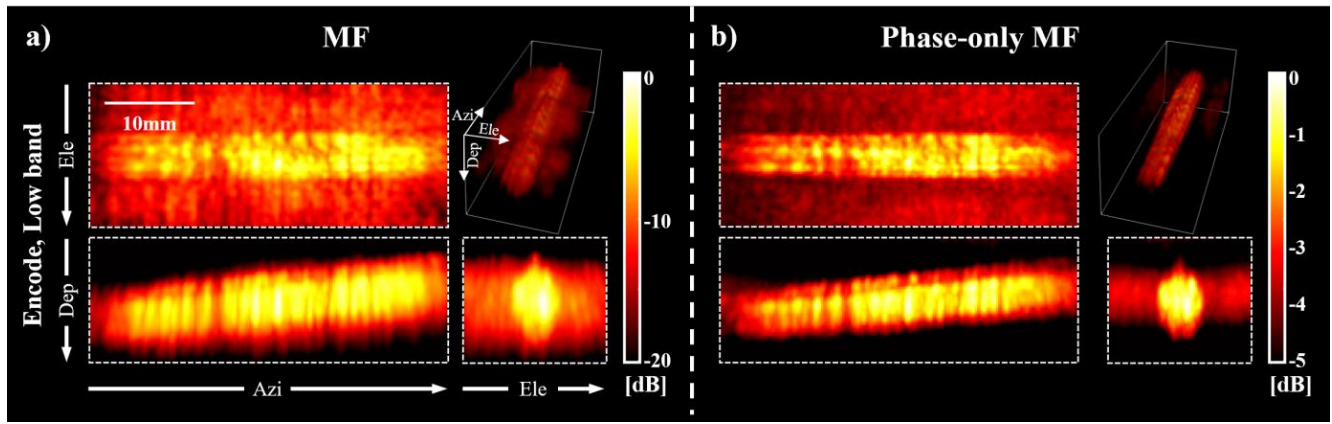

Fig. S2 Demonstration of phase-only MF in the phantom experiment for the encoded low-band case, shown as 3D PDIs with MIP and 3D rendering views. a) Standard MF reconstruction; b) Phase-only MF reconstruction. Rendering views are shown with 15 dB (left) and 5 dB (right) dynamic ranges. The phase-only MF yields a more complete flow profile with sharper boundaries, albeit with a reduced dynamic range.

**Other Supplementary Materials:**

Movie M1: Comparison of 4D PDI reconstructions in the flow phantom using **non-encoded** MF. Left: low-band; Right: high-band. Both reconstructions were obtained from the same acquisition and share the same ground truth.

Movie M2: Comparison of 4D PDI reconstructions in the flow phantom using **encoded** MF. Left: low-band; Right: high-band. Both reconstructions were obtained from the same acquisition and share the same ground truth.

Movie M3: 3D PDI rendering of the healthy volunteer's CA reconstruction, shown in rotational view.

Movie M4: Doppler spectrograms by depth (prior to clutter filtering) across three cardiac cycles from the in vivo measurement.